\documentclass[twocolumn,showpacs,amssymb,prd]{revtex4}

\usepackage{graphicx}
\usepackage{dcolumn}
\usepackage{bm}
\usepackage{epsfig}

\begin{document}
\newcommand{\gsim}{\hbox{\rlap{$^>$}$_\sim$}}
\newcommand{\lsim}{\hbox{\rlap{$^<$}$_\sim$}}

\title{Hyperluminal Signatures in the Afterglows \\
       of  Gamma-Ray Bursts 980425 and 030329}

\author{Shlomo Dado$^1$, Arnon Dar$^1$, A. De R\'ujula$^{2,3}$}
\affiliation{1) Physics Department, Technion, Haifa, Israel\\
2) Theory Division, CERN, Geneva, Switzerland \\
3) IFT, Universidad Aut\'onoma, Madrid, Spain}

\begin{abstract} 
The late-time high-resolution X-ray and radio 
observations of GRB980425/SN1998bw, the closest known gamma ray burst 
(GRB) associated with a supernova (SN) explosion, may have 
resolved the hyperluminal source that produced the GRB and its afterglow. 
Its hyperluminal speed $\sim 350\, c$ is consistent with that 
expected in the cannonball (CB) model of GRBs. The observed superluminal 
expansion of the late-time radio image of GRB030329/SN2003dh, 
the GRB with the brightest and longest followed up  radio afterglow  
to date, is also consistent with that expected in the CB model of GRBs and it
extrapolates to an apparent early-time hyperluminal motion.
\end {abstract}


\maketitle

The large cosmological distances of long duration GRBs 
were first indicated by observations with the Compton Gamma Ray 
Observatory (CGRO) shortly after its launch [1]. Combined with the very short 
rise-time of the GRB pulses it implied an implausibly large energy release in 
gamma-rays from a very small volume, if the emission was isotropic, as 
assumed in the original fireball models of cosmological GRBs [2]. A simple 
solution of this ``energy crisis" was that GRBs are not isotropic, but are 
narrowly beamed along the direction of motion of the highly relativistic 
jets, which produce them [3]. 
It was also suggested [3] that such jets are a succession of plasmoids of ordinary matter, 
similar to those 
launched in Galactic microquasars [4], but with much larger initial bulk 
motion Lorentz factor $\gamma(0)$. Presumably, they are launched in mass 
accretion episodes onto the newly formed compact object in stripped-envelope 
supernova explosions, in merger of compact stars (neutron stars, quark stars, 
black holes) in close binaries, and in phase transition of compact stars 
following cooling, loss of angular momentum, or mass accretion episodes [5]. 
It was further suggested that the GRB pulses are produced by inverse Compton 
scattering (ICS) of external ambient light by the electrons in the plasmoids [3].

The ``smoking gun" signatures of the ICS production mechanism of GRBs by such 
highly relativistic narrow jets are [3] a very large polarization, $P\approx 
2\theta^2\gamma^2/(1+\theta^4\gamma^4)\sim 100\%$, and early-time hyperluminal 
velocity (very large apparent superluminal velocity, $V_{app}\sim 
\gamma(0)\,c/(1+z)\gg100 $) in the plane of the sky for the most probable 
viewing angle $\theta\approx 1/\gamma(0)$ of GRBs at large redshifts. But, 
unlike the gamma ray-polarization, the hyperluminal velocity during 
the GRB pulses seemed at the time impossible to detect, because of the short 
duration, the large cosmological distances, and the poor localization of GRBs. 
The discovery of GRB afterglow [6] at longer wavelengths, which continues long 
after the GRB, opened the possibility to detect the large  superluminal 
velocities of the decelerating jets that allegedly produce the narrowly beamed GRB 
afterglows [7,8].

The radio afterglow of GRB980425, the closest GRB with a known 
redshift ($z\!=\!0.0085$), and the radio afterglow of the relatively 
nearby GRB030329 ($z\!=\!0.1685$), the brightest radio afterglow of a GRB 
detected to date, have offered the best opportunities to look for the 
hyperluminal signature [9] of the highly relativistic narrow jets, which 
according to the cannonball (CB) model of GRBs  produce both the GRBs 
[3] and their afterglows [7]. To the best of our knowledge, this has been 
overlooked in the late-time high resolution X-ray and radio 
observations of SN1998bw/GRB980425 [10].

In the case of SN2003dh/GRB030329, the observed late-time 
superluminal expansion of their radio image [11] indicates a hyperluminal 
expansion when extrapolated to early 
time ($t\ll 1$ day). These observations were 
compared [11] to predictions of the CB model of GRBs [12], which at the 
time were based on the assumption of jet-deceleration by elastic 
scattering of the medium in front of it [13] (``collisionless 
shock'' in the standard fireball models). This assumption did not reproduce 
well enough the observations of [11] and was
replaced in the CB model a decade ago with the assumption of CB 
deceleration by plastic collision with the medium. The model
was extensively confronted with 
the mounting observational data on GRBs and their afterglows.
It did not only successfully reproduce  the main 
properties and the detailed light-curves of GRB pulses and GRB afterglows, but
its major falsifiable predictions (those which do not depend on free 
adjustable parameters and/or multiple choices) were confirmed by many 
observations [14]. The only exception so far, has been claimed in [11]. 

In this letter, we present an updated analysis of the late-time X-ray and 
radio observations of SN1998bw/GRB980425 [10] and SN2003dh/GRB030329 
[11], which provides evidence --consistent with that predicted by the CB 
model-- for a hyperluminal motion of the source of the afterglows of 
GRB980425 and GRB030329 relative to their parent supernovae.
 
{\bf Superluminal motion of CBs:}
The velocity in the plane of the sky of a light-source
moving at redshift $z$ with a velocity  $v=\beta\, c$ at 
an angle $\theta$ relative to the line of sight is given by [15] 
\begin{equation}
V_{app}=\beta\, \gamma\,\delta\,sin\theta\, c/(1+z)\,,  
\label{Eq1}
\end{equation}
where $\gamma=1/\sqrt{1-\beta^2}$ is its bulk motion Lorentz factor,
and $\delta=1/\gamma(1-\beta\cos\theta)$ is its Doppler factor.
For a given $\beta$, 
$V_{app}$  as a function of $\theta$  has a maximum  
$\beta\,\delta \,c/(1+z)$ at $\sin\theta=1/\gamma$.

If the light source is a highly relativistic CB with $\gamma^2\gg 1$ 
and $\theta^2\ll 1$, its Doppler factor is well approximated by 
$\delta\approx 2\,\gamma/(1+\gamma^2\,\theta^2)$, and Eq.(1) becomes
\begin{equation} 
V_{app}\approx 2\,\gamma^2\,\theta\, c/(1+\gamma^2\,\theta^2)(1+z)\,.
\label{Eq2}
\end{equation}
Ordinary GRBs are produced by highly relativistic 
CBs with a viewing angle that satisfies  
$\gamma_0\,\theta\approx 1$ where $\gamma_0=\gamma(t=0)$. Consequently, 
they have a hyperluminal speed $V_{app}\sim \gamma\, c/(1+z)$,
while CBs viewed far off axis, i.e., CBs 
with  $\gamma_0^2\,\theta^2\gg 1$, have  
\begin{equation}
V_{app}\approx 2\,c/\theta\,(1+z)\,.
\label{Eq3}
\end{equation}
Deceleration of such CBs decreases their $\gamma$
and reduces their $V_{app}$.  For $\gamma\approx 1/\theta$, 
\begin{equation}
V_{app}\approx c/\theta (1+z)
\label{Eq4}
\end{equation}
and when $\gamma^2\,\theta^2\ll 1$, 
\begin{equation}
V_{app}\approx 2\,\gamma^2\,\theta\,c/(1+z)\,.
\label{Eq5}
\end{equation}

\noindent 
{\bf Hyperluminal signature in GRB980425?} Kouveliotou et al. 
[10] reported X-ray observations of the environs of SN1998bw/GRB980425 
on day $t1=1281$ post GRB with the Chandra X-Ray Observatory (CXO). The 
observations resolved two declining point-like sources, S1a and S1b. The 
location of S1a coincided with the radio location of SN1998bw. The second 
point-like source S1b was located $\Delta\theta_1=2.04"\pm 0.30"$ away 
from SN1998bw. It was suggested to be an ultraluminous X-ray (ULX) 
transient at a much larger redshift [10].

A weak radio source was detected by Soderberg et al. [10] at 1.7" west 
and 3.1" north of the position of SN1998bw with the 
Australia Telescope Compact Array (ATCA) on day $t2=2049.19$ post 
GRB980425. The authors pointed out that the source was too bright to be 
due to thermal emission from the complex of HII regions in this area 
[10], and that it could be the radio counterpart of the transient X-ray 
source (S1b) detected by Kouveliotou et al. [10]. However, in that case, 
the radio location of S1b 2049.19 days after GRB980425 at 
$\Delta\theta_2=3.54"$ from SN1998bw moved $1.5"\!\pm 0.30"$ farther away 
from its sky location on day 1281.

If S1b was launched by SN1998bw, 
and  2049.19 days later its radio emission was  observed  
at $\Delta\theta_2=3.54"$,  
its average velocity in the plane of the sky  must have been
\begin{equation}
\langle V_{app}\rangle=D_L(bw)\,\Delta\theta_2 /t_2  \approx  355\, c,      
\label{Eq6} 
\end{equation}
where $D_L(bw)=35.6$ Mpc is the angular distance to SN1998bw, assuming, 
as hereafter, a standard cosmology with $H_0 =71$ km/s/Mpc, 
$\Omega_M=0.27$, and $\Omega_\Lambda=0.73$. 
By day 1281, such a source should have reached an angular 
displacement of $\Delta\theta\approx (1281/2049.19)\, 3.54"=2.21"$ from 
SN1998bw, 
consistent with the observed displacement $\Delta\theta_1=2.04"\pm 0.30"$ of S1b. 

Could S1b be the ``hyperluminal" CB that produced the single-pulse 
GRB980425 and its afterglow?

In the CB model, GRBs and their afterglows are produced by CBs with 
$\gamma\gg 1$ and $\theta\ll 1$. Ordinary GRBs are viewed from small angles 
($\theta^2\,\gamma_0^2\, \lsim 1$), while low-luminosity GRBs and XRFs are 
ordinary GRBs viewed from larger angles, $\theta^2\,\gamma_0^2\gg 1$. 
This interpretation of low-luminosity GRBs and XRFs [9,13] is based 
on the fact that while ordinary GRBs, low-luminosity GRBs and XRFs 
all appear to be produced by broad-line stripped-envelope SNeIc very similar 
to SN1998bw, their isotropic equivalent energies $E_{iso}$ 
are orders of magnitude smaller than those of ordinary GRBs. For example, 
SN1998bw, which produced the low-luminosity GRB 980425 [16] with 
$E_{iso}\approx 8\times 10^{47}$ erg [17] was very similar to SN2013cq 
[18], which produced the very bright GRB130427 with $E_{iso}\approx 
8.5\times 10^{53}$ erg.
Moreover, the short lifetime of the massive stars that die as SNeIc and the SN-GRB/XRF association  yield observed rates of GRBs, low-luminosity GRBs and XRFs that are all proportional with the same proportionality constant to the star formation rate, when their viewing-angles and the
detection threshold are properly taken  into account [19].

In the CB model, GRBs are produced by inverse Compton scattering of glory 
light --the light halo formed around the progenitor star by scattered 
light from pre-supernova ejections-- by the electrons enclosed in the CBs. 
The low-luminosity GRB980425 was a single-pulse GRB
[20]. In the CB model, the pulse was produced by a single CB viewed far 
off axis [9,13]. Its superluminal velocity, given by 
Eq.(3), depends only on its viewing angle, as long as 
$\gamma^2\,\theta^2\gg 1$. This  angle can be 
estimated in three ways: from the peak energy $E_p$ of the time-integrated 
spectrum of the single pulse of GRB980425 in the ``GRB rest frame", 
from the duration of this pulse, and from the light 
curve of the X-ray afterglow.

The peak energy $E'_p$ 
of the time integrated spectrum of a GRB pulse in the rest-frame of the 
exploding star (indicated by a prime) is given by [13]
$E'_p=\gamma_0\,\delta_0 \langle\epsilon_\gamma\rangle$  
where $\langle\epsilon_\gamma\rangle$ is the mean energy of the 
photons of the glory 
surrounding the exploding star. If GRBs are viewed from different angles 
but their CBs and glory are otherwise 
 approximate standard candles 
(of similar mean $\langle\epsilon_\gamma\rangle$ and $\gamma_0$)
the product $\gamma_0\,\theta$ in the far off-axis GRB980425 satisfies
\begin{equation}
\gamma_0^2\,\theta^2\approx 2\,\langle E'_p\rangle/E'_p\,, 
\label{Eq7}   
\end{equation}
where $E'_p=55\pm 15$ keV is the peak energy of GRB980425 [17],
and  $\langle E'_p\rangle $ is  the mean peak energy of ordinary GRBs,
viewed from an angle $\theta\sim 1/\gamma_0$. The  
85 ordinary GRBs ($E_{iso}>10^{52}$ erg) with measured $E'_p$ 
listed in [17], have a mean value $\langle E'_p\rangle=563\pm 114$ keV. 
Hence, Eq.~(7) yields  $\gamma_0\,\theta\approx 4.5\pm 0.7$.
This value and the CB model best estimate,  
$\langle \gamma_0\rangle \approx 700$, from
analyses of  hundreds of GRBs [19], imply
that the viewing angle of GRB980425 was
$\theta\approx (6.5\pm 1.0)\times 10^{-3}$. 

An estimate based on the measured 30-500 keV light curve of the single 
pulse GRB980425 [20] yields a similar value of $\theta$: In the CB model, 
the observed full width at half maximum, $\Delta t$, of a single 
pulse in a GRB is proportional to $\gamma\,\delta/(1+z)$. Consequently,
\begin{equation} 
\gamma_0^2\,\theta^2\approx (\Delta t/1.0085)/(\langle \Delta t 
/(1+z)\rangle)\,. 
\label{Eq8} 
\end{equation} 
The single pulse of GRB980425 in the energy band 30-500 keV, 
measured with BATSE aboard 
the CGRO, 
had $\Delta t\approx 13$s [20]. The resolved single pulses of ordinary 
GRBs with known redshift, have 
$\langle \Delta t/(1+z)\rangle\approx 2.2$s
in the Swift energy band. Consequently, for GRB980425 [21], Eq.~(8) yields 
$\gamma_0\,\theta \approx  3.44$ and $\theta\approx 4.9\times 10^{-3}$.

The CB model fit to the 0.3-10 keV X-ray light-curve of GRB980425 
shown in Fig.1 yields $\gamma_0\,\theta \approx  3.54$.
\begin{figure}[]
\centering
\epsfig{file=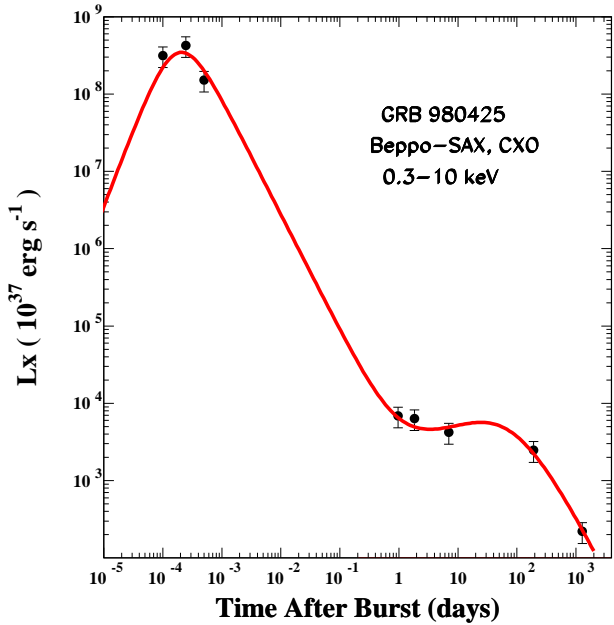,width=9.0cm,height=9.0cm}
\caption{
The 0.3-10 keV X-ray light-curve of GRB980425/ SN1998bw 
measured by Beppo-SAX [10] (first 7 points). The last point at 1281
days corresponds to the source S1b resolved by CXO [10]. 
The line is the CB model best fit light-curve of the prompt single 
pulse and  afterglow of GRB980425.}
\label{fig1}
\end{figure}  

The  above three estimates yield a mean value $\theta\approx 5.5\times 
10^{-3}$. If the source S1b was launched in the SN1998bw explosion and 
maintained $\gamma^2\,\theta^2\gg1$ until day 2049.19, its superluminal 
velocity $V_{app}\approx 355\,c$ must have satisfied Eq.(4), which implies 
$\theta=5.7\times 10^{-3}$ consistent with the 
above three independent CB model estimates.
But, did the CB satisfy the condition $\gamma^2\theta^2\gg1$?

The path-length of a highly relativistic CB of  baryon number $N_b$ and 
radius $R$ that decelerates from an initial $\gamma_0^2\theta^2 \gg1$ to 
$\gamma^2\theta^2=1$ by sweeping in the matter of an external  medium
with a constant baryon density $n$, is:
\begin{equation}
L\approx 4\, \gamma_0^2\, t_d\,c\,[\gamma_0^2\,\theta^2-1]\,,    
\label{Eq9}
\end{equation}  
where $t_d=(1+z)N_b/8\,\pi\, n\,R^2\,c\,\gamma_0^3$. The values 
$\gamma_0\,\theta=3.54$ and $t_d=0.28\,d$ obtained from the best fit CB 
model light-curve (Fig.1) to the X-ray light-curve of GRB980425, and 
$\gamma_0=700$, yield $L\approx 5$ kpc. 
This is $\sim 100$ times larger than the thickness of the 
disk of the face-on dwarf galaxy ESO 184-G82E, the host of GRB980425 
[23]. Therefore, we conclude that the CB,  launched in SN1998bw 
and generating GRB980425 and its afterglow, still retained 
$\gamma_0^2\theta^2\gg1$ on days 1281 and 2049.19 post-GRB when it moved 
through the halo of its host galaxy into the intergalactic space. Hence, 
its superluminal speed was $V_{app}\approx (2/\theta)\,c\approx 
351\,c$, consistent with $V_{app}\approx 353\,c$ of the S1b 
source observed by Chandra [10] in  X-rays on day 1281, 
and by ATCA [10] in radio on day 2049.19.

\noindent
{\bf Hyperluminal signature in GRB030329.} 
The relative proximity ($z$=0.1685) of GRB030329 and its record-bright 
radio afterglow made possible its record long, high resolution follow-up 
observations with the Very Long Baseline Array (VLBA) and the Global Very 
Long Baseline Interferometry (VLBI) Array, until 3018.2 days post GRB [11].
Assuming a disk shape, the radio image of GRB030329/SN2003dh 
was fit with a circular Gaussian of diameter $2\,R_\perp(t)$ 
yielding a mean apparent expansion [11],
\begin{equation} 
\langle\beta_{app}\rangle=2\,R_\perp/c\,t\,\propto t^{-1/2}
\label{Eq10} 
\end{equation}

SN2003dh and GRB030329 had individual image sizes much smaller than the 
resolution of the VLBA and VLBI arrays: the initial large expansion 
velocity of broad-line SNeIc, such as SN2003dh, decreases to less than 
10,000 km/s within the first month, beyond which it continues to decrease, 
roughly like $t^{-1/2}$. Such an expansion velocity yields SN image-sizes 
$<0.002$ pc and $<0.005$ pc on days 25 and 83 after burst, compared to the 
joint image-size of GRB030329/SN2003dh, $\sim 0.2$ pc and $\sim 0.5$ pc, 
respectively, extracted from the radio observations [11]. As for CBs, time 
dilation and ram pressure suppress their lateral expansion in the 
circumburst medium, as long as they move with a highly relativistic speed. 
Although their small size implies radio scintillations, the large 
time-aberration washes them out completely in the time-integrated 
measurement in the observer frame -- the typical $dt\sim 100$ minutes 
integration time of the VLBA observations [11] corresponds to $dt'=\gamma 
\delta\, dt/(1+z)$, i.e., many years, in the GRB/SN rest frame!

The VLBA and VLBI measurements [11] could not resolve such small images of 
SN2003dh and the CB it ejected, nor a superluminal motion of the CB. 
However, if the size of their joint radio image as measured in [11] is 
adopted as a rough estimate of the time-dependent distance between the 
source of the GRB afterglow (the CB) and the SN, it can be used to test the 
CB model, as shown below.

In ordinary GRBs, CB deceleration in an ISM with a constant density 
yields the late-time behaviors $\delta(t)=2\,\gamma(t)\propto t^{-1/4}$, 
$F_\nu\propto t^{-\alpha_\nu}\,\nu^{-\beta_\nu}$ with 
$\alpha_\nu=\beta_\nu+1/2$, $V_{app}= 2\,\gamma^2\,\theta\, c 
/(1+z)\propto t^{-1/2}$, and a mean expansion rate in the interval $(0,t)$
$\langle V_{app}(t)\rangle \approx 2\, V_{app}(0)$.  These behaviors
do not depend on any adjustable parameters and are well satisfied by the 
late-time X-ray afterglow [25] of GRB030329. E.g., its measured 
spectral index $\beta_x=1.17\pm 0.04$ in the 0.2-10 keV X-ray band [25], 
yields a late-time temporal decay index $\alpha_x=\beta_x+1/2=1.67\pm 
0.04$, in agreement with the observed $\alpha_x=1.67$ [25], as shown 
in Fig.2. 
\begin{figure}[]
\centering
\epsfig{file=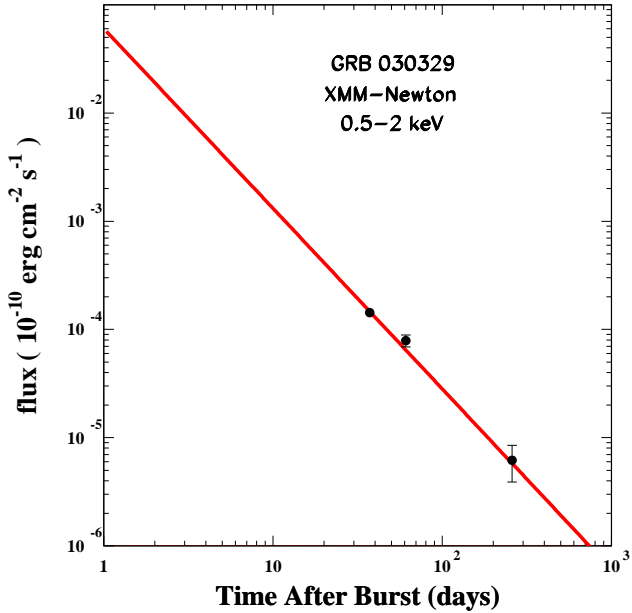,width=9.0cm,height=9.0cm}
\caption{The late-time 0.5-2 keV X-ray afterglow of the joint source 
GRB030329/SN2003dh as measured by XMM-Newton [25]. The line is the CB 
model  with its predicted
$\alpha_x=\beta_x+1/2$,  with the measured $\beta_x=1.17\pm 0.04$ [25].}
\label{Fig2}
\end{figure}

The late time VLBA and VLBI radio measurements of the image-size of 
GRB030329/SN2003dh are also in good agreement with the CB model 
prediction $\langle\beta_{app}(t)\rangle\propto t^{-1/2}$ for the 
GRB-SN separation as
long as the CB has not reached the halo of the host galaxy, as  
shown in Fig.3. The CB model prediction is for $\gamma_0\, \theta=1.915$ 
obtained from the width of the first $\gamma$-ray pulse of GRB030329, 
$\gamma_0=\langle \gamma_0\rangle\approx 700$ of ordinary GRBs, and a 
deceleration 
parameter $t_d\approx 8\times 10^{-5}\,d$. At early time 
$\langle\beta_{app}(t)\rangle$ does not depend on $t_d$. Beyond the 
afterglow break, its dependence on $t_d$ is rather weak, $\propto 
t_d^{1/2}$ as long as the CB moves in constant density medium. 
The canonical values of the CB model parameters yield
$t_d\approx 5.2\times 10^{-5}$, for a canonical ISM with $n=1\, {\rm cm}^{-3}$,
close to its best fit value, suggesting deceleration 
in an edge-on host galaxy.
\begin{figure}[]
\centering
\epsfig{file=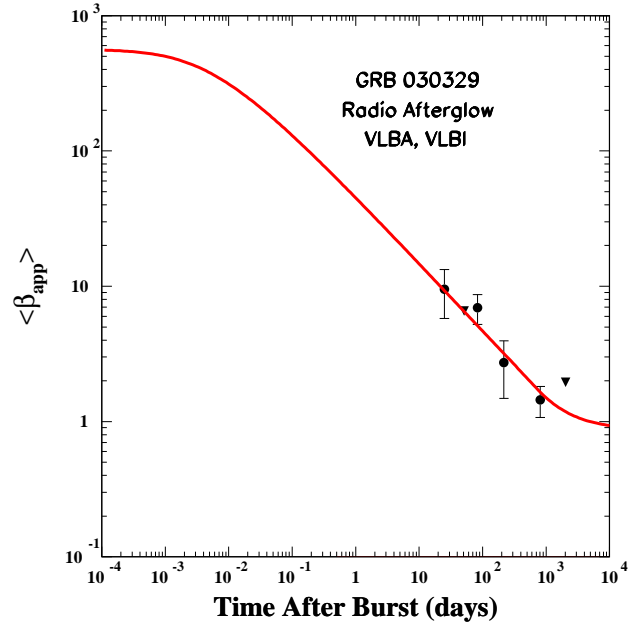,width=9.0cm,height=9.0cm}
\caption{The time-averaged expansion rate of the radio image of 
GRB030329/SN2003dh [11]. The line is the predicted $\langle 
\beta_{app}\rangle$ of the CB,  taking $2R_\perp$ in Eq.(10) 
as an estimate of its distance  
from SN2003dh.}
\label{fig3}
\end{figure}

{\bf Conclusions:} The late-time high-resolution follow-up observations 
of the X-ray and radio afterglows of GRBs 980425 and 030329 suggest that 
the jets which produce GRBs in broad-line SNeIc akin to SN1998bw and 
SN2003dh have initially a hyperluminal velocity in the plane of the sky 
($\beta_{app}\gg 100$), as in the CB model of GRBs. Whether they maintain 
it at late time depends on the density along their direction of motion in 
the host galaxy. The observed late-time behavior $2R_\perp\propto 
t^{1/2}$ of the radio image size of GRB030329/SN2003dh, is that predicted 
by the CB model for a CB moving away from SN2003dh in a constant density 
ISM. The simple extrapolation of the observed late-time behavior $\langle 
\beta_{app}\rangle\propto t^{-0.5}$ [11] to $t\sim 100$ s, the beginning 
of its afterglow, yields a hyperluminal velocity $>500$ c [26].

If the source S1b near SN1998bw observed by CXO on day 1281 is the CB
which produced GRB980425 and its afterglow,  it must have had a hyperluminal
velocity $\sim 350 c$, and must have moved by day 2049 to roughly the location of the
radio source seen by ATCA on day 2049. By now it must be $\sim 11"$ away from
SN1998bw. It may still have there a faint but detectable radio afterglow. If,
contrariwise, S1b and the ATCA source are unrelated to GRB980425, 
they  may still be seen unmoved from their positions on day 1281, and 2049,
respectively.


\begin{thebibliography}{999}
\bibitem{Comment21}
In the CB model, $E_{iso}$ is proportional to the photon column density 
encountered by the jet of CBs, launched in the SN explosion.  
This column density  depends on the pre-supernova activity. Lack 
of knowledge of the photon column density impedes a precise 
prediction  of the GRB viewing angle from the measured value of 
$E_{iso}$.
\end{thebibliography}
\end{document}